\documentclass[5p,twocolumn,authoryear]{elsarticle}

\usepackage{natbib}
\usepackage[fixlanguage]{babelbib}
\usepackage[english]{babel}
\usepackage{color}
\usepackage{url}

\usepackage{graphicx}

\usepackage{amssymb}

\newcommand\astrogridd{{AstroGrid\mbox{-}D}}
\newcommand\dgrid{{D\mbox{-}Grid}}

\begin{document}
\begin{frontmatter}

\title{AstroGrid-D: Grid Technology for Astronomical Science}

\author[AIP]{Harry Enke}
\author[AIP]{Matthias Steinmetz}
\author[MPA]{Hans-Martin Adorf}
\author[AEI]{Alexander Beck-Ratzka}
\author[AIP]{Frank Breitling}
\author[ZAH]{Thomas Br\"usemeister}
\author[MPE]{Arthur Carlson}
\author[MPA]{Torsten Ensslin}
\author[ZIB]{Mikael H\"ogqvist}
\author[AIP]{Iliya Nickelt}
\author[AEI]{Thomas Radke}
\author[ZIB]{Alexander Reinefeld }
\author[TUM]{Angelika Reiser}
\author[TUM]{Tobias Scholl}
\author[ZAH,NAOC]{Rainer Spurzem}
\author[ZAH]{J\"urgen Steinacker}
\author[MPE]{Wolfgang Voges}
\author[ZAH]{Joachim Wambsgan\ss}
\author[AIP]{Steve White}

\address[AIP]{Astrophysikalisches Institut Potsdam AIP, Potsdam, Germany}
\address[MPA]{Max-Planck-Institut f\"ur Astrophysik MPA, Garching, Germany}
\address[AEI]{Max-Planck-Institut f\"ur Gravitationsphysik (Albert-Einstein 
Institut) AEI, Potsdam, Germany}
\address[ZAH]{Astronomisches Recheninstitut am Zentrum f\"ur Astronomie 
Heidelberg ZAH, Heidelberg, Germany}
\address[ZIB]{Konrad-Zuse-Zentrum f\"ur Informationstechnik Berlin ZIB, Berlin, 
Germany}
\address[MPE]{Max-Planck-Institut f\"ur extraterrestrische Physik MPE, Garching, 
Germany}
\address[TUM]{Technische Universit\"at M\"unchen, Institut f\"ur Informatik TUM, 
Garching, Germany}
\address[NAOC]{ National Astronomical Observatories of China, Chinese Academy of Sciences
 NAOC/CAS, 20A Datun Rd., Chaoyang District, Beijing 100012, China}

\begin{abstract}
We present status and results of \astrogridd{}, a joint effort of 
astrophysicists and computer scientists to employ grid technology for 
scientific applications.
\astrogridd{} provides access to a network of distributed machines with a set
of commands as well as software interfaces. It allows simple use of
computer and storage facilities and to schedule or monitor compute tasks and 
data management. It is based on the \emph{Globus Toolkit middleware} (GT4).

Chapter 1 describes the context which led to the demand for advanced software solutions
in Astrophysics, and we state the goals of the project. 

We then present characteristic astrophysical applications that 
have been implemented on \astrogridd{} in chapter 2. We describe simulations of
different complexity, compute-intensive calculations running on multiple sites (\ref{Compute}), 
and advanced applications 
for specific scientific purposes (\ref{Advanced}), such as a connection to robotic telescopes 
(\ref{Rob}). We can show from these examples how grid execution improves 
e.g. the scientific workflow.

Chapter 3 explains the software tools and services that we adapted or newly 
developed. Section \ref{Working} is focused on  
the administrative aspects of the infrastructure, to manage users and 
monitor activity.
Section \ref{Components} characterises the central components of our architecture:
The \astrogridd{} information service to collect and store metadata, a 
file management system, the data management system, and
a job manager for automatic submission of compute tasks.

We summarise the successfully established infrastructure in chapter 4, concluding 
with our future plans to establish \astrogridd{} as a platform of modern e-Astronomy.

\end{abstract}


\end{frontmatter}

\section{Introduction}
\label{Intro}

Astrophysical research is an intent driver for advances
in computer science, especially so for high performance computing and data 
intensive calculations. We are used to the continuous increase of processor power
which increases the potential of computer 
based analysis.
Even faster is the rise of sensor size and storage capacity, both of which 
in recent years have grown even stronger than \emph{Moore's Law} would predict. 
Unfortunately, this trend of growing data volumes also increases the complexity of the
data management, as well as the processing, analysis and visualisation. Above a 
certain level, new methods have to be applied, e.g. the management of data becomes
a task that is no longer trivial enough for a file system alone.
This challenge affects many other domains outside of 
Astrophysics in the same way, and it is an important challenge
to find answers, since in several research areas further progress depends on
the successful processing of data volumes in the high Terabyte or Petabyte scale. 

One solution for improved data management is the recent success in meta data 
standardisation and advanced corresponding protocols. In astrophysics this
approach has led to the international "Virtual Observatory" initiative, which
now allows for a fast search within extensive volumes of diverse stored
data.	

But computer science itself has also researched ways to improve infrastructure
usage and simplify the processing of information. 
The most compelling answer of recent years was the 
massive development in Grid computing, where a new software layer is used to 
connect distributed information infrastructures like clusters, storage servers 
and desktops to a loose network (see: \citep{ITF09}).

Several research grid infrastructures 
were successfully set up in the past years. The most impressive example is the US "TeraGrid", 
funded since 2001 by the National Science Foundation. It offers over a petaflop of total 
compute capabilities and many different services and gateways to thousands of US scientists. 
Like the Open Science Grid, TeraGrid is based on the Globus Toolkit, enlarged
by an auxilary software package set.

The European enterprise \emph{EGEE} ("Enabling Grids for E-SciencE") was started 
2004 as a EU project, sponsored from the European Union's research framework.
EGEE was at the beginning mostly driven by the CERN's new large Hadron Collider and its demand 
for compute power. It currently combines about 40.000 CPUs and will in 2010 be transferred 
into a new body called EGI (European Grid Initiative). It will then focus mostly on the role 
to coordinate the collaboration of the national grid initiatives with supported middlewares 
limited to gLite, UNICORE and ARC.

The German national Grid initiative was inaugurated in 2004 by the Federal Research ministry. 
It has seen two main stages: \dgrid{} 1 (2005-2008) focussed on Grid application for 
fundamental sciences, whereas \dgrid{} 2 (2007-2010) mostly researched Grid use in applied sciences 
and industry. 

The \astrogridd{} project was part of the first \dgrid{} initiative and started 
in 2005. Five major German astronomy institutes participated: AIP, AEI, MPA, 
MPE, and ZAH, together with computer science groups from the ZIB Supercomputer
center and TUM. They collaborated on the common project goal: To establish 
a collaborative working environment for astronomy which provides the users 
with the powerful and reliable software tools and allows easy access to 
compute and storage facilities for their scientific work. 

To achieve this the projected aimed to:
\begin{itemize}
\item set up a grid-based infrastructure for astronomical
  and astrophysical research
\item embed existing computational facilities, astronomical software applications, 
data archives and instruments
\item integrate this grid infrastructure into the national \dgrid{} environment
\item provide support for other astronomical groups to join  
\item strengthen international partnerships 
\end{itemize}

\astrogridd{} has reached these goals in its setup phase which ended early 2009.
The most important results were the first Virtual Organisation management, 
now the \dgrid{}-standard (see \ref{VOM}), integration of
special hardware \dgrid{} (\ref{Nbo}, \ref{Rob}) and the production run of
one of the most compute intensive scientific grid application to date (\ref{Geo}).

We hereby present our experiences and results in some detail. The paper
is grouped into two main chapters: First the astrophysical applications 
(\ref{Use}) and secondly our developments in information technology (\ref{Services}).
In the summary (\ref{Summary}) we give an outlook on our future plans.

\section{Astronomy and Grid: Astronomical Use Cases running on the
project network}
\label{Use}

Most areas of Astronomical research can profit from
e-Science concepts and grid technology in particular. 

In the course of the project, a total of twenty selected astronomic pilot 
applications were modified for grid use and implemented.
Use cases ranged from
compute-intensive simulations running on clusters, task-farming 
jobs to explore large parameter spaces, analyzing programs accessing 
astronomical databases, to complex and specific applications as
described below.
These \textit{use cases} also served to define the requirements for 
\astrogridd{} components.

When considering a grid implementation for a given application, it
is decisive to compare how time-consuming and complex
the task will be compared to the benefits, such as speed gain. 
Before we describe examples in detail we will state general experiences 
for different application classes.

For \emph{large simulations}, e.g. from cosmology
{\it (Mare Nostrum, \cite{URL})}, a grid environment is ideal to reduce typical obstacles.
In a grid infrastructure, a unified and standardised interface
is provided to access the grid-enabled resources of a high performance 
computing center. The Grid offers a common way to 
execute calculations and manage resulting data.
Also many details, such as efficient data transfer, are handled by
the Grid middleware. The need to learn details about a specific
center is minimised.

{\em Taskfarming jobs}
benefit from the grid infrastructure since there now is a multitude of 
resources available to them, as shown for the Geo600-example (\ref{Geo}).
Especially applications with limited requirements can gain immensely from a grid 
implementation, where many hundreds of instances can be executed concurrently.

{\em Robotic telescopes} (\ref{Rob}) serve as an example for special scientific 
hardware. When combined to a worldwide 
network on the basis of grid middleware, this brings important advantages to 
coordinated observations. 
Typical tasks for such a network are multi-wavelength campaigns or 
the continuous monitoring of transient astronomical objects. 
A grid based network simplifies
coordination and infrastructure management, since grid devices such as storage
servers and databases are easy to connect. Moreover, global grid schedulers can
automatically coordinate and optimise the observations.

For {\em large data sets} like the Sloan Digital Sky Survey {\it (SDSS, \cite{URL})} 
or the Millenium simulation archive \citep{2005Natur.435..629S}, 
efficient processing poses a huge
problem. The data often have inconsistent formats
and interfaces, and the methods still vary how to define subsets and
correlate them, or even run algorithms against them.
To select data, the scientist needs access to a given database and, in most cases, 
also access to additional data files. Corresponding results must be stored in 
some accessible device. Since the data volumes are growing large
and the catalogues may be distributed, techniques for data discovery 
searching, and transmission (data streaming) are applied, 
combined with mechanisms for parallelisation and load-balancing 
for the computing processes.
At this point, Grid data processing overcomes the limits of the centralised
data processing approach where so far large volumes of data are 
transferred to the application that requests them. 
The alternative is to distribute the data processing
within the grid and the use of storage facilities accessible via grid methods.
Whenever possible the application is executed at the location of the data.

Many solutions and design decisions, such as described in the last paragraph, 
rely on the 
work and standards of the Virtual Observatory. 
Hence \astrogridd{} collaborates closely 
with the German Astrophysical Virtual Observatory (GAVO), for example 
when using GAVOs easy-to-use data access interface to $N$-body simulations. 
Via GAVO's participation in the IVOA activities,
\astrogridd{} also participated from the developments 
where grid middleware is used to provide VObs services. 
We will continue the collaboration between \astrogridd{} and GAVO  
in the creation of a virtual data center for astronomy.

To support users in the deployment of their application, we compiled an 
\emph{application-to-grid} guide that illustrates the steps
to grid-enable simple applications {\it (App2Grid, \cite{URL})}.
\subsection{Compute-Intensive Generic Applications}
\label{Compute}

Many compute-intensive applications can be subdivided into
multiple small parallel tasks that can run independently, 
e.g. on multiple grid resources.
This can usually be achieved by partitioning the physical properties
of the relevant parameter space. In the following, we will discuss three 
such compute-intensive grid applications, namely the task-farming 
use case Dynamo, NBDOY6++ as an example use case with little I/O, 
and the gravitation wave analysis tool GEO600. 

We have found that a grid implementation for this application type 
can be very beneficial and achieved within a manageable timeframe.

\subsubsection{Dynamo}
\label{Dyn}
The \emph{Dynamo} package shows how to use the 
advantages of grid computing without complex programming. 
Grid implementation is achieved by a shell script, 
that is lean, relatively simple to understand and easy to configure. 
It provides a grid connection
for the purpose of \emph{task farming} of serial programs, i.e. the
launching of many 
instances of scientific software where the input differs for each run.
We call this type of application \emph{atomic}, since as a serial
calculation it requires no further communication until the results are
produced.

The scientific problem for this example is derived from the field of
Magneto-Hydro-Dynamics. Rotation and turbulence in stars,
accretion disks, and galaxies produce a magnetic field by the dynamo 
effect. In the
case shown here the numerical simulation solves the induction
equation with a turbulent electromotive force (alpha tensor).
The general parameter dependence as well as the time
development of a given set are studied, with special focus on the
``flip-flop''-phenomenon of star spots \citep[see][]{dynamo_elstner2005}.

\begin{figure}[!ht]
  \centering
  \includegraphics[width=\linewidth]{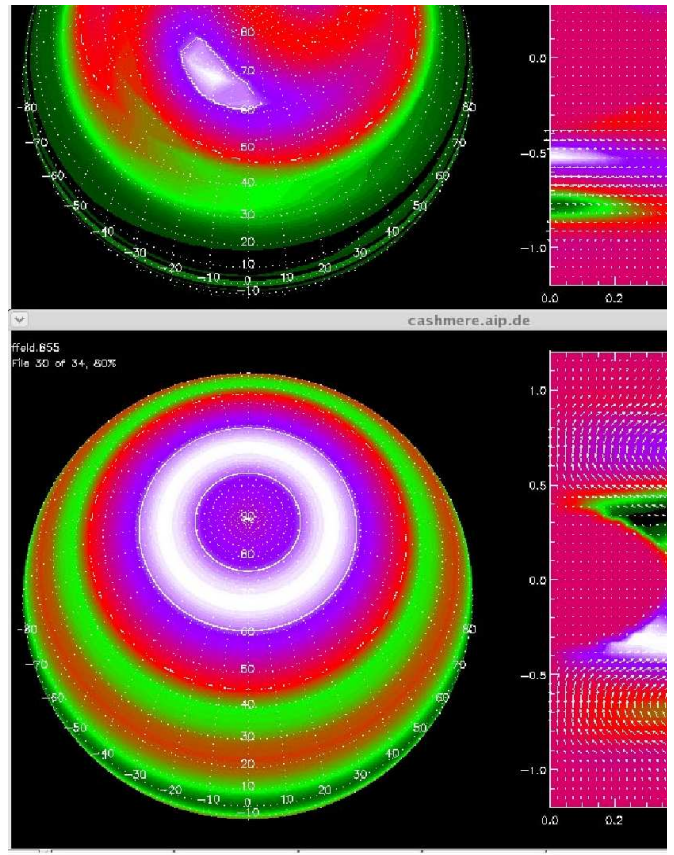}
  \caption{Example output of a Dynamo run, showing real time results of four 
  different grid resources}
  \label{fig:dynamo-output}
\end{figure}

For grid task farming with varying input sets the script reads in any
number of input directories, each of which contains different data.
Together with the executable, the job is then submitted iteratively to
grid resources specified in a list and executed there.
Intermediate output can be retrieved on the fly;
a visualisation example is shown in Fig.~\ref{fig:dynamo-output}.
 
This solution is currently being applied to a similar use case for
GAVO.
Upgrades of the software would properly include GridSphere
and improve the stage-in process. 
The script package can be downloaded from the \astrogridd{} use case 
web pages {\it (Dynamo, \cite{URL})}. Users with a demand for atomic, serial jobs 
should find this solution easy to implement within \astrogridd{} or
within similar, Globus-based grids. 
\subsubsection{NBODY6++ and {$\varphi$}GRAPE}
\label{Nbo}
NBODY6++ and {$\varphi$}GRAPE are two variants of a family of high-order
accurate direct $N$-body simulation codes, which are built upon the
development of a series of earlier versions (1-6) of NBODY codes 
\citep{Aarseth1999}. {$\varphi$}GRAPE is the only parallel code of
this type to use special purpose GRAPE6 hardware \citep{Harfst2007}
based on  GRAPE which has 
been designed by the University of Tokyo to accelerate gravitational force 
computations between particles \citep{Makino2003, Fukushige2005}. 
While {$\varphi$}GRAPE
is just a plain direct parallel NBODY code using a 4th order Hermite integrator 
with
hierarchical block time steps, NBODY6++ is a parallel version of NBODY6 (with 
regularisation
of close encounters, Ahmad-Cohen neighbour scheme, and other features), which is
optimised for parallel general purpose supercomputers \citep{Spurzem1999}.

Examples for applications where gravitational forces between many bodies have to 
be calculated are globular clusters,
young forming star clusters or central dense star clusters in galactic nuclei.
Recent typical research using direct $N$-body simulations includes,
e.g., models of 
galactic star clusters with many binaries \citep{Hurley2007} or
massive binary black holes embedded in dense stellar systems leading to 
coalescence
and gravitational wave emission \citep{Berczik2005, Berczik2006, Berentzen2009}.

NBODY6++ and {$\varphi$}GRAPE are use cases of \astrogridd{} which supports 
deployment and execution of these as jobs on its resources using single and 
parallel hardware,
as well as parallel hardware with special purpose GRAPE cards.
The ZAH offers the 32 node GRACE cluster {\it (GRACE, \cite{URL})} as a resource
of \astrogridd{}, with
reconfigurable specialised hardware to a total peak speed of 4 Teraflop/s \citep{Harfst2007, 
Spurzem2007, Spurzem2008}.
Another resource with GRAPE hardware integrated in the \astrogridd{} is a cluster 
at the Main Astronomical Observatory in Kiev, Ukraine {\it (MAOKIEV, \cite{URL})}, also
an example of collaboration made possible on the basis of a grid Virtual
Organisation.

Submission of an NBODY job starts with a shell script preparing
an XML-based job description
which is then staged and transported
through the \astrogridd{} Globus middleware. Input data, output data and files
go along with the job submission process. Future goals are to allow the
submission of NBODY jobs through a portlet under the \astrogridd{} web portal and
an integration of the \astrogridd{} file management system to allow handling of
large datasets independent of the job staging process, see deployment 
instructions and tutorial {\it (NBODY6++, \cite{URL})}.

\subsubsection{GEO600}
\label{Geo}
The GEO600 use case is a task farming
application. It uses the \textit{Einstein@Home} application
for analysing the data of the GEO600 Laser Interferometer near Hannover,
in order to find signals of gravitational
waves. 

Einstein@Home is an ideal candidate for a grid
application because of multi-platform support, well tested software
base, simple resource requirements, built-in checkpoint and recovery
methods, adjustable run time, and linear scaling with node number.
Within the \astrogridd{} project we developed the software 
for grid deployment, job statistics and the details for 
constant production mode runs, such as restart 
after a regular job end and cleanup of recoverable errors. 

The deployment is triggered by a script which is invoked in a 
Web Service Grid Resource Allocation and Management (WS--GRAM)
job to all grid machines on which the GEO600 jobs should run. 
As prerequisites on the target resource only
Subversion (to retrieve the GEO600 source code) and a Perl
interpreter are necessary. All other required
software is installed during the deployment.

Depending on the number of currently pending and active tasks, the submission 
script will automatically determine when to submit new tasks to a grid
resource. To establish a continuous submission scheme it is
therefore sufficient to invoke the script periodically on the
target.
\begin{figure}[ht]
  \centering
  \includegraphics[width=\linewidth]{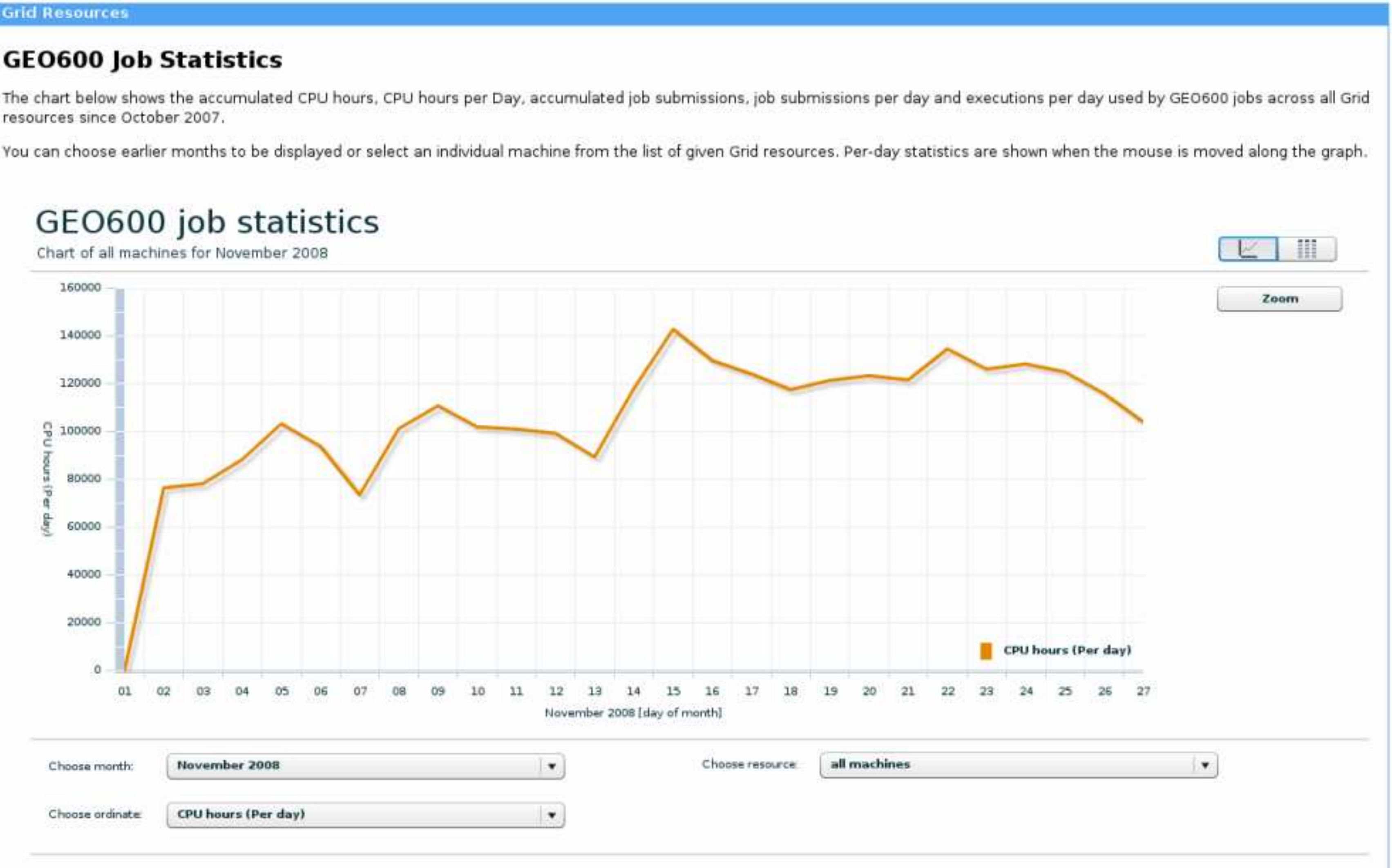}
  \caption{GEO600 CPU Time November 2008, taken from {\it (GEO600-statistics, \cite{URL})}
   with x-axis days of month, y-axis CPU hours consumed (sum over all used Grid resources)}
  \label{fig:GEO600Stats}
\end{figure}

The intermediate data are stored on the execution hosts since
a central server approach would significantly slow down the job 
transfer rates. The submission of the GEO600 jobs can be controlled 
from a single workstation, from which the execution hosts are contacted 
directly. We plan to use the \astrogridd{} scheduler Gridway for the
distributions of the GEO600 jobs in a future update. 
Furthermore it is foreseen to extend the GEO600 use case to
grids based on other grid middlewares than Globus, such as gLite or 
Uniform Interface to Computing Resources (Unicore). This would
allow a further distribution of the Einstein@Home jobs in the grids
available.

The GEO600 use case has been running in production mode for more than a year, 
and it consumes around 100\hspace*{1mm}000 CPU hours a day on \dgrid{} resources 
(see Fig.~\ref{fig:GEO600Stats}).
\subsection{Advanced Applications}
\label{Advanced}
Special purpose astrophysical applications and complex tool environments 
can also benefit from a grid infrastructure. We have chosen four relatively 
different use  cases to represent this class of 
astrophysical applications and to show how we approach an implementation.

First, {\em Clusterfinder} is a use case involving
both the deployment and performance of a typical compute-intense
data analysis application and the extensive use of distributed data 
resources.
{\em Cactus} shows how monitoring and steering 
methods for parallel numerical simulations in the grid can be generalised, 
and how a web portal can provide user-friendly 
assessment of grid jobs and visualisation.
Access to {\em robotic telescopes} as a grid resource represents a unique approach
to a grid with heterogeneous elements.
Finally, the Planck Process Coordinator Workflow Engine 
{\em ProC} has been grid enabled to demonstrate the power of grid computing
when applied to the complex workflow of processing the data product of a 
satellite mission. It is a useful example for the handling of observation 
which may exceed the local
capabilities  or must be organised to suit the demands of a locally distributed 
working group.

\subsubsection{Clusterfinder}
\label{Clu}
Clusterfinder is an example for the deployment 
of a compute-intense astrophysical application that uses distributed
data, and its increase in performance.
The scientific purpose of Clusterfinder is to reliably identify clusters
of galaxies. It correlates the signature of X-ray images with that in catalogues 
of optical observations in order to study the large scale
structure of the universe.
Scanning at optical wavelengths to look for areas
with an unusually large number of galaxies is not an
unambiguous method to identify large clusters, as the galaxies may
be spread out along the line of sight.
Also the observation of the X-ray emission of the hot gas between 
galaxies will result in some false identifications as 
there are many other X-ray sources.
In order to combine both sources of information, the theory of point processes is applied 
to calculate the statistical likelihood of a cluster at any point in space,
and peaks in the combined likelihood are extracted into a catalogue of
galaxy clusters {\it (Clusterfinder, \cite{URL})}.

Data retrieval and the calculations can easily be parallelised as
the algorithm for any point in the sky depends only on data from 
nearby points, making Clusterfinder well-suited for grid implementation. 
Input of the Clusterfinder program consists of a cosmology and galaxy cluster
model, together with the grid of sky coordinates and redshifts on which the 
likelihood is to be calculated. 
Scanning the available data consumes about 20,000 CPU-
hours per model. This entails over two years on a single processor or only
several days when the resources of \astrogridd{} and \dgrid{} are used. 
An exploratory calculation on a smaller area can be executed on the grid in one night.

To implement Clusterfinder for a grid environment, two software tools were developed:
A "grid-module" handles the installation and compilation on the resource, and an 
"environment" suite ensures that the necessary files and connections are available
on any resource.

The logistics of performing Clusterfinder calculations on the grid
involves
splitting the calculation into jobs that can run in parallel,
identifying grid hosts with the capacity to accept a job at the given
time, reassembling the individual results into a coherent whole, and
documenting the internal and external conditions under which the
calculation was carried out. A single calculation is then submitted 
as a globus job and calculates a 
likelihood map with a given set of parameters. 
The results are collected using either the 
post-staging capabilities of Globus or by direct grid transfer using 
the globus-url-copy command.
In the case of Clusterfinder, special consideration has been given to the
input data. The SDSS and ROSAT all sky survey 
{\it (RASS, \cite{URL})} catalogues are too large to copy the 
complete data set to a grid node. Therefore the 
makefile controlling the Clusterfinder workflow is set up to request 
just the data needed from these catalogues.

A demonstration version of Clusterfinder is available as a portal
application. The user can input coordinates and retrieve the corresponding likelihood
map. It is planned to extend this portal to provide a production version
of Clusterfinder as a grid service, including control over all the input
parameters and even the files for the cosmological model. 
\subsubsection{Cactus}
\label{Cac}
The {\em Cactus Computational ToolKit (CCTK)} {\it (Cactus, \cite{URL})} is an open
source, general purpose software framework designed to solve large-scale
systems of partial differential equations on supercomputers using finite
differencing techniques.
In the Astrophysics science community Cactus is used to numerically
simulate extremely massive bodies, such as neutron stars and black holes, and
analyse the gravitational wave signal patterns emitted by these objects
as predicted by Einstein's theory of General Relativity.

In \astrogridd{} we have developed application-specific techniques for Cactus
which enable scientists to manage their simulations more efficiently and in a
more collaborative context. 
Many of these methods make use
of standard grid technology internally {\it (Deliv. 6.6, \cite{URL})}.

As an example for online application monitoring and steering, users can connect
to a running Cactus simulation just like any standard secure
Hypertext Transfer Protocol (HTTP) web service, with a browser of their favorite 
choice.
User authentication and authorisation is based on X.509 grid certificates
(see Section \ref{VOM}).
When logged in, users can query an up-to-date status of the
simulation (e.g. the physical simulation time or
{\tt stdout/stderr} log output).
Built-in online visualisation methods are available
to analyse intermediate simulation data graphically via dynamic generation
of 1D line or 2D surface plots, thus allowing users to evaluate the quality of
the simulation while the application is still running.
Once authorised, they can also steer the simulation by interactively changing 
parameters, triggering a checkpoint to be written, or by terminating the job gracefully.

Each Cactus simulation submitted to some supercomputer or grid resource
can also announce itself at startup to the \astrogridd{} information service,
by sending an RDF document with metadata uniquely describing the simulation.
The information service is then able to keep a history of all simulations
submitted by Cactus users. To access and search that simulation database we 
provide
a Cactus portlet, based on {\em GridSphere} (see Section \ref{Inter}) as a 
standardised web interface.
After logging into the portal, users can query the list of Cactus runs
and filter it by owner, execution host, specific parameter settings
etc. Queries are implemented as Cactus-specific GridSphere portlets 
{\it (Deliv. 7.5, \cite{URL})}, 
allowing the user to easily navigate through the list of simulations and browse 
individual query results.
Also available in the portal are the results of nightly Cactus integration
tests, which are performed automatically on various machines in the grid,
in order to verify the correctness of the latest development version of the 
code.
\subsubsection{Robotic Telescopes}
\label{Rob}
In recent years a growing number of ground-based robotic telescopes have been 
comissioned in
astronomy, due to their increased technical reliability. Robotic astronomy 
allows observations from sites which may be astronomically favourable, 
but are otherwise remote or even hostile for human
operators, e.g. Antarctica.

With more robotic telescopes becoming operational,
there has been increasing interest in interconnecting them.
Such a telescope network can accomplish new types of obervations. 
Examples are an uninterrupted observational campaign 
over many hours independent of day time and weather as required in
astro-seismology, and rapid multi-wavelength observations in case of
transient events.

\astrogridd{} contributes to this development with
the {\it (OpenTel, \cite{URL})} software package.
OpenTel achieves the integration of robotic telescopes into the \astrogridd{}
infrastructure and implements a telescope network based on grid middleware. 
Each telescope thus acts as an individual grid resource with its own 
grid certificate. One immediate 
advantage provided by grid technology is the direct connection to compute and
storage resources for data analysis and archiving. Additionally, grid user and
virtual organisation management provides a good solution for the central 
management of access rights.

The metadata management relies on Stellaris
(cf. section \ref{Inf}) and the {\it (Usage Record format \cite{URL})} of the Global
Grid Forum transformed into RDF. The metadata is retrieved from Stellaris
using Simple Protocol and RDF Query Language (SPARQL) queries. The monitoring of 
observations is similar to the
observation of jobs described below in section \ref{Mon}.
The \textit{Robotic Telescope Markup
Language} (RTML) \citep{RTML} of the \textit{Heterogeneous Telescope Network}
(HTN) \citep{HTN} serves as the protocol for observation requests.

The \textit{OpenTel Tools} package provides programs for the tasks of observation
(job) submission, cancellation, and status queries. The programs are based on 
commands of the Globus Toolkit and are executed from the command line. 
Further details are described in {\it (Deliv 5.3, \cite{URL})} and in the package documentation.

\begin{figure}[th]
  \centering
  \includegraphics[width=\linewidth]{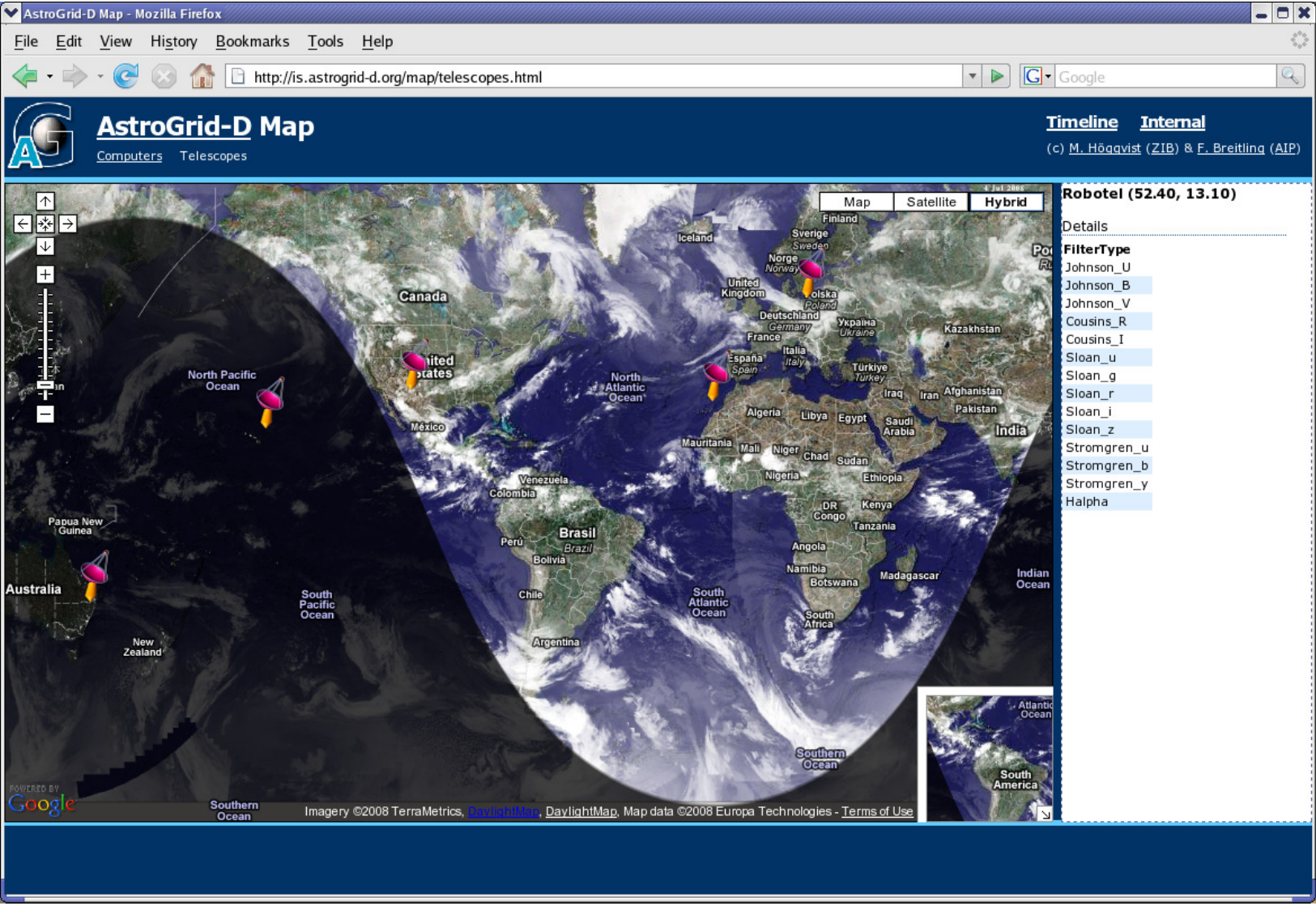}
  \caption{The Telescope Map is an interactive user interface for the selection
    of telescopes. Daytime, weather conditions as well as the geographic
    location and properties of the telescopes are displayed.}
  \label{fig:TelescopeMap}
\end{figure}

Several user interfaces have been
developed to simplify operation management: 
the \textit{OpenTel Tools}, the \textit{Telescope Map}, the
\textit{Telescope Timeline}, a broker, and a scheduler.
The \textit{Telescope Map} is an interactive user interface shown in Fig.
\ref{fig:TelescopeMap}. It is an extension of the \astrogridd{} \textit{Resource
Map} (section \ref{Mon}) for displaying geographic locations of telescopes
and their properties such as available filters. Also displayed
are day and night regions as well as weather information.

\begin{figure}[th]
  \centering
  \includegraphics[width=\linewidth]{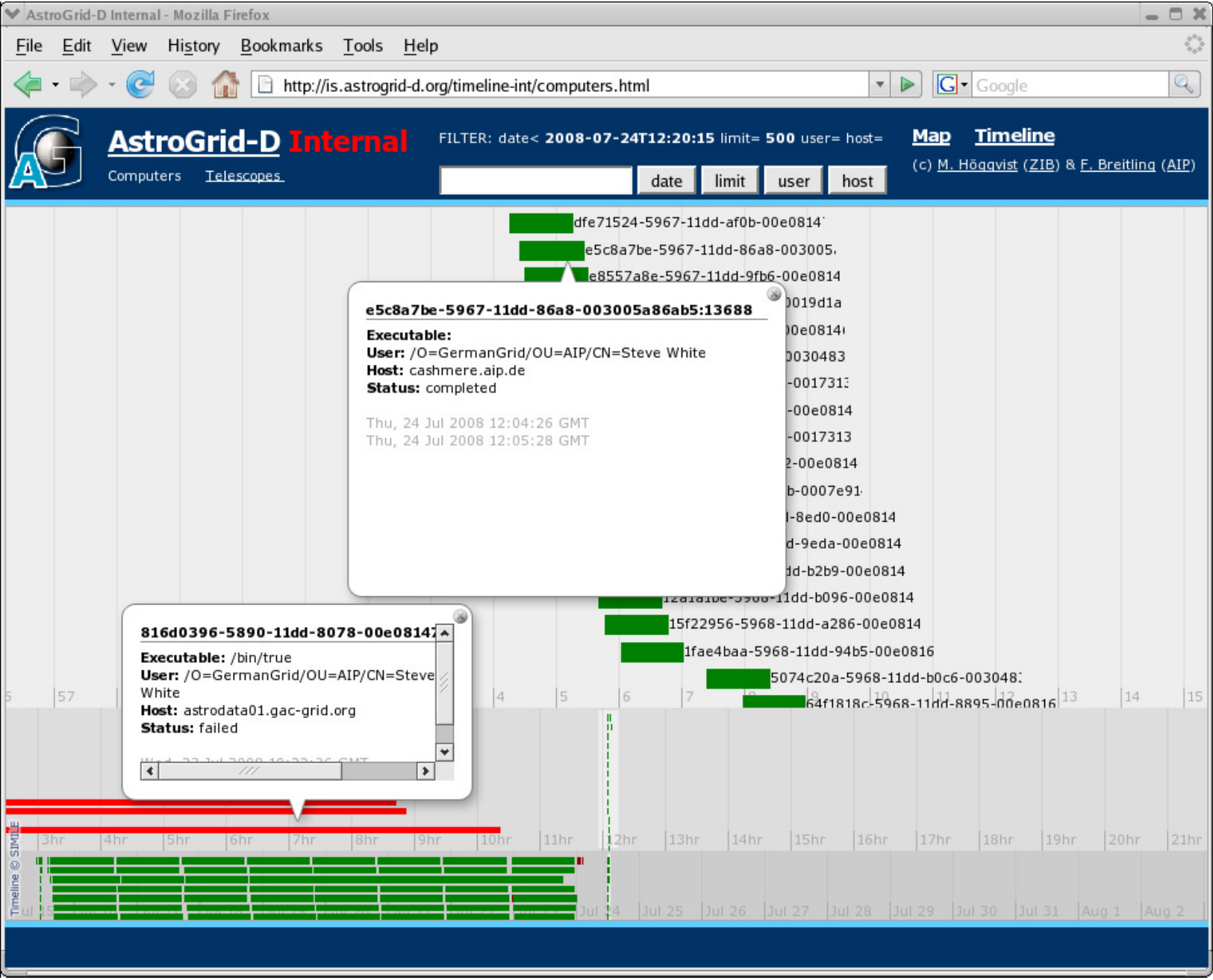}
  \caption{The Timeline is an interactive user interface for displaying status,
    progress and general information of grid jobs. The top area displays hours, 
	each line displaying the duration of a job and an identifyer. A mouseclick 
	opens up an information window (inset), displaying indepth information about the job. 
	In the below areas the scope of display is days and months. }
  \label{fig:JobTimeline}
\end{figure}

The \textit{Telescope Timeline} is another interactive user interface useful for
monitoring {\it (Deliv. 2.7, \cite{URL})}. It is an extension of the \astrogridd{} 
\textit{Timeline} (\ref{Mon}) and displays information about executed observations with an
appearance similar to Fig. \ref{fig:JobTimeline}.

The \textit{broker} achieves an automatic selection of telescopes based on the
requirements of an observation {\it (Deliv. 5.5, \cite{URL})}. Filters and geographic coordinates 
but also the
dynamic data such as the current weather conditions are examples for selection
criteria.

\begin{figure}[ht]
  \centering
  \includegraphics[width=\linewidth]{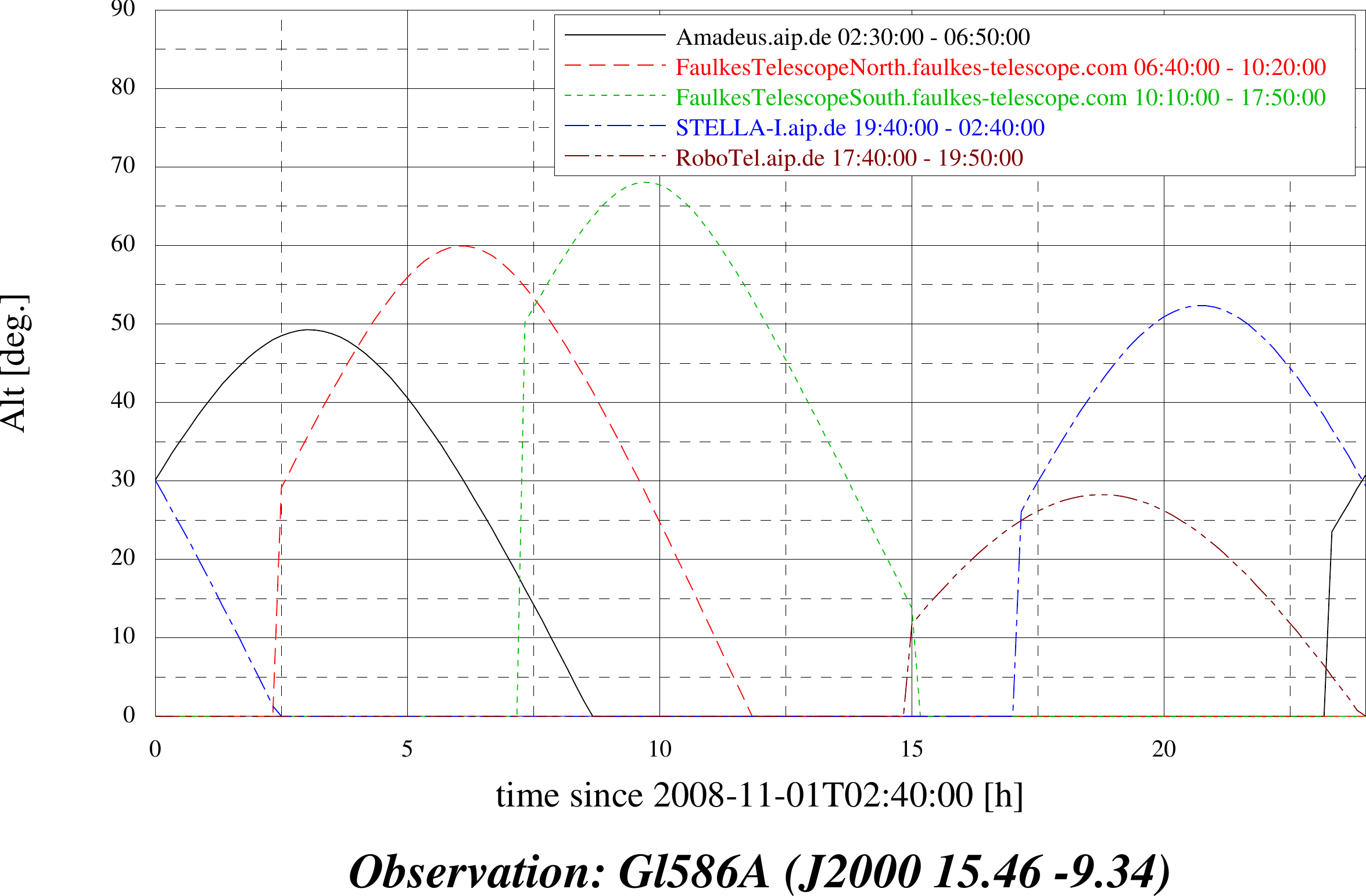}
  \caption{Altitudes versus observation time for the network schedule of a simulated 24\,
    h observation of Gl586A. The plot is produced by the
    OpenTel scheduler. The intersections of the altitude curves provide the
    time intervals for the observations by the different telescopes. 
    Schedules are optimised for object altitude.}
  \label{fig:Schedule}
\end{figure}

The \textit{network scheduler} generates observation schedules of the desired 
duration {\it (Deliv. 5.8, \cite{URL})}.
Whenever necessary, an observation is handed over to be continued by another 
telescope of the network. An example
for a 24\,h observation of the star Gliese 586A (Gl586A) in the small network of Fig.
\ref{fig:TelescopeMap} is shown in Fig. \ref{fig:Schedule}. 

The OpenTel software has been tested with the AIP's robotic telescope STELLA-I
\citep{STELLA} and simulated networks. It is available at {\it (OpenTel, \cite{URL})}.

\subsubsection{The ProC workflow engine for scientific grid-computing}
\label{Pro}
The Process Coordinator (ProC) is a scientific workflow engine. It was originally 
developed as an integral component of the software infrastructure for the Planck 
Surveyor satellite mission of the European Space Agency 
\citep{2000SPIE.4011....2B}.

\begin{figure}[!ht]
  \centering
  \includegraphics[width=\linewidth]{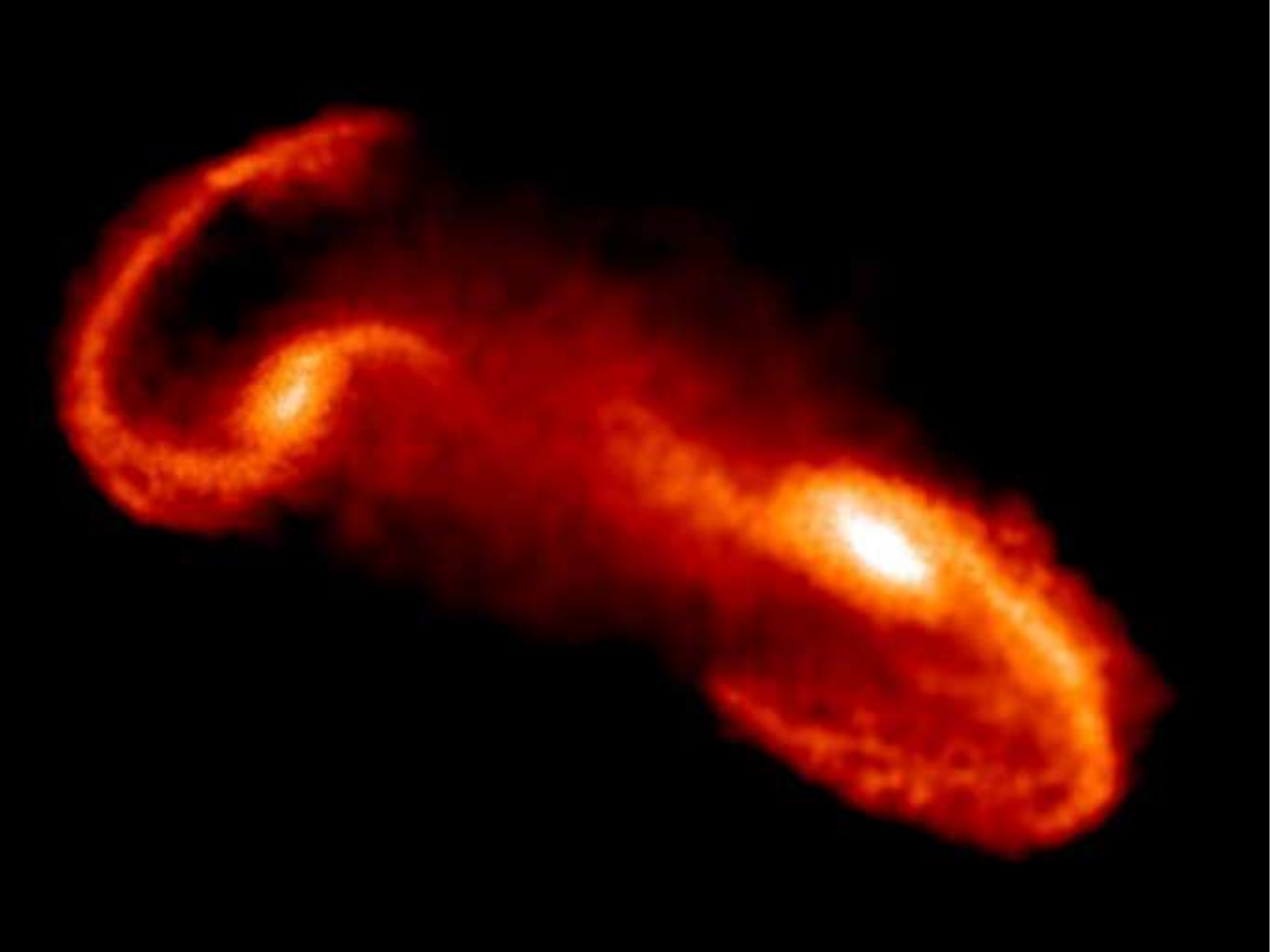}
 \caption{ProC supported simulation: Galaxy collision calculated with the 
GADGET-simulation package steered by the ProC sampling control element.}
 \label{fig:applications2}
\end{figure}

Currently, two sets of scientific programs are being executed using the ProC, 
each forming a problem-domain
specific toolbox. One is the simulation and data analysis package required for 
the Planck mission and cosmic
microwave background (CMB) research \citep{2006A&A...445..373R}. The other is a 
post-processing package
for GADGET-simulations of cosmic structure formation 
\citep{2005MNRAS.364.1105S}, shown in Fig.
\ref{fig:applications2}. Both cosmological
research areas are expected to benefit strongly from the parallel computing 
resources now being accessible
for parameter space sampling problems via the grid-enabled ProC.

The ProC software package consists of three components: a graphical workflow 
editor, 
a graphical user interface for workflow execution, and a workflow engine, 
equipped with
an application programing interface (API) and a versatile command line interface 
for expert users. The ProC is implemented platform independently in Java and 
uses the extensible markup language (XML).

One of the advantages of using the ProC, compared to simple scripting, is 
its ability to automatically
recognize opportunities for reusage of previously generated computational 
results and for parallel execution
of computational units. This latter capability can exploit multiple cores on a 
single processor, multiple
processors cooperating in a local cluster, or the hundreds of compute elements 
offered by a dispersed grid.

With the help of the ProC Pipeline Editor the user is able to compose and modify 
scientific workflows consisting of programs, data flows, and control elements of the ProC library. 
Strong data-typing assures that only valid connections between modules can be 
made.
The ProC's feature set includes typical control elements (e.g. loops), 
a fork/join mechanism, and specialised 
``sampling'' elements for the investigation
of high-dimensional parameter spaces via various algorithms. These elements 
permit user-controlled parallel
execution of the same program segment on different data. 

Within \astrogridd{} the ProC was grid-enabled with the help of the Grid Application 
Toolkit (GAT, cf. subsection \ref{subsecGAT}).
In sample runs we used 200 compute elements simultaneously on a remote grid node. 
The need to deploy non-portable scientific code to a large number of grid nodes 
entailed the development of a comprehensive package of environment modules.

Upon request the ProC package is available free of charge for scientific 
computing purposes.

\section{The \astrogridd{} Services}
\label{Services}
In this section we describe the architecture of our
grid implementation and explain the role of several of its components and
services.

We decided to base the astrophysical community grid 
on a recent version of Globus Toolkit (GT4) as a 
most widespread and advanced middleware solution.
However, grid middleware capabilities are only 
generic functions and need enhancements to be of actual use. 
In more general terms the middleware serves as an abstraction layer or
translation interface. It connects the \emph{resource} (the individual hardware 
and its operating system) with the \emph{grid 
resource API} (application programming interface) and with a set of uniform commands 
and applications, called the \emph{middleware API}. The last interface is 
the one presented to the grid users and grid applications. 
An operational grid thus in some ways resembles a 
nonlocal operating system with enhanced capabilities, such as distributed storage 
or access to connected clusters and their batch systems.

\begin{figure}[ht]
  \centering
  \includegraphics[width=\linewidth]{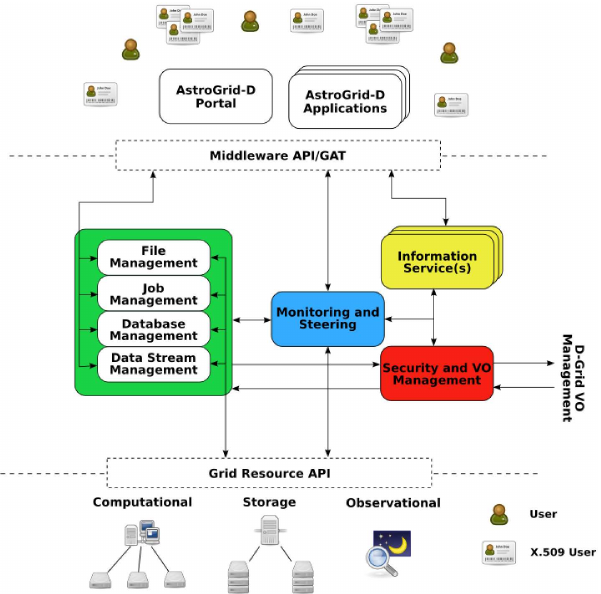}
  \caption{Sketch of \astrogridd{} architecture, showing the layers and services 
that are involved in a grid application, and the pathes of interaction.}
  \label{fig:architecture}
\end{figure}

In a second step we then modified or added architecture elements as necessary
for Astronomical applications.
The result is shown in Fig.~\ref{fig:architecture}.
At the resource level we find compute elements (CE), storage elements (SE), and instruments. 
While compute and storage elements are common to all grids and can be properly managed by the basic 
middleware,
the inclusion of instruments (e.g. robotic telescopes) is one of the additions made by 
\astrogridd{}. Another addition is \astrogridd{}'s central information service Stellaris (\ref{Inf}) 
which stores metadata of components, services and data (yellow block in Fig.~\ref{fig:architecture}).

We further extended the middleware \emph{capabilities} for job and file management (green block in Fig.~\ref{fig:architecture})
by adding data stream management (\ref{Data}). Other components were enhanced: Monitoring 
and steering were attached to the Stellaris information service (blue block in Fig.~\ref{fig:architecture}). 
With our Virtual Organisation management we achieved user and group management 
based on the GT4 security layer (red block in Fig.~\ref{fig:architecture}), to implement a grid that can easily be used by
collaborations to share access rights and data.

Fig.~\ref{fig:architecture} only illustrates the architecture components.
Not shown is the underlying, interconnecting network and the security layer.

\subsection{Working with \astrogridd{} Resources}
\label{Working}
Each system with built-in security requires the user and even services (hosts, 
databases etc.) to authenticate themselves. 
The grid uses X509-certificates, i.e. public/private key encryption,\label{X509} for this purpose. 
At least one Grid Certification Authority per country
provides such certificates for users, resources and services. With this certificate it is possible 
to log onto other grid resources from any grid enabled workstation.  

The following subsections
describe some details of grid and resource work.
The subsection about \emph{VOrg Management} shows how the collaborative concept of 
virtual organisations and the security layer are tied together. Then, a brief overview 
of the procedures for 
integrating a resource into the grid is provided.
The last paragraphs introduce different interfaces 
provided by \astrogridd{}.

\subsubsection{VO Management}
\label{VOM}
Virtual Organisations (VOrgs, often somewhat confusingly called \emph{VO}'s) 
are a central element of any grid. In some
aspects they are the grid representation of the more familiar ``group'' concept 
of an operating system. A VOrg is formed by any number of users with a
common intention to share resources, data and access rights in a grid.

In \astrogridd{} any user is authenticated 
by an individual X.509 certificate. However, the certificate itself does not allow 
access to resources of \astrogridd{} or \dgrid{}, 
since that right is restricted to members of our 
main VOrg ``\astrogridd{}''. Thus each user must also register for membership to 
this VOrg.

To improve the registration process and administer the members, \astrogridd{} 
uses a service written
by Fermilab, the \emph{Virtual Organisation Membership Registration Service}, 
{\it (VOMRS, \cite{URL})}. The
registration service itself is only accessible with the user's certificate 
installed in the web browser. During the registration process some
of the user's work details are collected, such as name and institution. 
The user also has to choose which of the available VOrgs he wants to 
belong to. Upon verification by the user's institute, 
the VOrg Administrator will grant the membership status. 

Additionally to the 
main VOrg, in \astrogridd{} 
currently four smaller VOrgs exist. These 
Sub-Organisations are used by specific institutes for internal grids, 
for our robotic telescope resources and our collaboration with GAVO.

To connect the VOrg member database of \astrogridd{} with each resource, we 
developed a separate 
service. At each resource this service regularly queries the central VOrg 
database for changes, and the resulting user list is applied to the resource's 
local access management. When an accepted VOrg member then logs on to the resource and 
is properly authenticated by the Globus Toolkit, he is mapped to an individual, 
local UNIX user account.

Our extension to the VOMRS offers a number of options for local resource 
administrators, e.g.
to import only specific VOrgs or white- or blacklist single users. The system 
also supports OGSA-DAI (see Section \ref{Data}) and Unicore user formats and cluster
options. Individuals who change their
``distinguished name'' string, e.g. due to a change of institution, can be 
mapped back to their former grid account. Even if there is in general no guarantee 
for user data to persist in the grid, it is often
useful to re-gain an existing environment of local settings and libraries.

\astrogridd{} established the VORMS based solution in 2006. Since then it was 
stable in operation, managing
the about 100 users of \astrogridd{}.
The successful concept was then also adopted by the German \dgrid{} 
where it became the standard form of user management.
\subsubsection{Resource integration}
\label{Res}
\astrogridd{} is currently comprised of about 20 grid resources provided
by its member institutions: computer clusters, workstations,
data storage servers, as well as a telescope server. 

German astronomers apply for inclusion of a computer resource into
\astrogridd{} on an individual basis; all German academic institutions
are eligible by default. Resources of an Ukrainian institution
have also been included for collaboration.

Ideally, to bring a resource onto the grid takes about fifteen hours for
an experienced administrator. In practice more time may be required, 
due to complications in networking,
retrieving certificates, and operating system peculiarities.
Why would an institute invest that work and put their valuable computer resources on
the grid? First, there is anyway considerable overhead for sharing of 
resources between institutions: accounts have to be set up, ports opened for 
special communications, etc. These problems are solved by bringing 
resources onto the grid and using the tools and standard solutions it provides.  
Second, on the grid, a resource
has a much wider group of users and can be used to full potential.

All steps required to bring hosts on-line as \astrogridd{} resources are 
described at {\it (AGD-Globus, \cite{URL})}.
\subsubsection{Monitoring}
\label{Mon}
In a distributed, diverse grid environment, the monitoring of its parts and processes 
is of central importance for users and administrators. 
Monitoring can in principle be divided in two categories: resource
and job monitoring.

Resource monitoring for compute and storage resources is realised in
\astrogridd{} through the \textit{Monitoring and Discovery System} (MDS) of the
Globus Toolkit. MDS is a suite of web services to monitor and discover
resources and services on Grids. The gathered information is displayed on
the \astrogridd{} resources overview web page {\it (MDS, \cite{URL})}.
An independent monitoring mechanism has been developed for
telescope resources, which handles telescope-specific information 
such as weather.

As a complementary interface to the resource list view, \astrogridd{} 
has developed a \textit{resource map} as an advanced user interface for displaying collected
resource information topographically. The \textit{Telescope Map}
in Fig. \ref{fig:TelescopeMap}, discussed in Section \ref{Rob},  
is a specialisation for telescopes. Both are based on the Google maps API.
When a resource is selected, additional information about its load and usage
is displayed. The information is obtained via SPARQL queries from {\it (Stellaris, \cite{URL})},
after it has been extracted from MDS, converted into RDF and uploaded to
Stellaris.
The Resource Map can be accessed at {\it AGResourceMap, \citep{URL})}. The
software can be obtained from the \astrogridd{} web page.

Job monitoring is based on globus' audit logging. Audit logging writes job
status information into a database. This information is translated into RDF/XML 
and transferred to Stellaris.

The \astrogridd{} \textit{timeline} was developed as a plain user interface to job
information. It is based on the {\it (simile timeline
\cite{URL})}. Jobs are represented by horizontal lines of length
proportional to the job duration. A colour code represents the status. For each
job, additional information such as user ID and name of executable can be
displayed. The search for information can be limited with keywords and in the public area 
the details are strongly reduced for privacy reasons.

Further details about monitoring can be found in {\it (Deliv. 5.9, \cite{URL})}.
\subsubsection{User and Developer Interfaces}
\label{Inter}
In \astrogridd{} there are four different ways available for actual grid use.
The middleware itself provides a commandline interface as well as an API for software.
The Grid Application Tool (GAT) provides an alternative API which hides the underlying grid 
middleware and makes its use 
transparent. And finally, GridSphere enables developers to quickly 
develop portlets for grid applications. Both GAT and GridSphere
do not require the installation of a grid middleware on the
submission host, and it is also possible to use them on Windows machines.\\

{\it The Globus Commandline and API} \nopagebreak

The \astrogridd{} resources are grid enabled by Globus middleware. They can thus be
accessed via the command line interface of Globus.
This 
interface allows data transfers and submission of jobs to the
grid and provides many more operations. 
For applications, Globus offers a rich API for each component of the 
middleware. 

The {\it Grid Application Toolkit} \nopagebreak
\label{subsecGAT}
{\it (GAT, \cite{URL})} is an API which offers grid access irrespective of the
middleware which connects the resource to the grid. The GAT Engine and
preliminary adaptors have been developed as part of the EU funded {\it ({Gridlab,} \cite{URL})}. 
Within the \astrogridd{} project the Java implementation of
JavaGAT is used. \astrogridd{} added adaptors for SGE, PBS, WS--GRAM
and gLite, and recently also a UNICORE adaptor
(UNICORE 6) was contributed by the DGI--2 project. JavaGAT currently
features adaptors to all the grid middlewares, which are used in \dgrid{}.
JavaGAT uses the security layers of the middleware. 
\begin{figure}[ht]
  \centering
  \includegraphics[width=.5\textwidth]{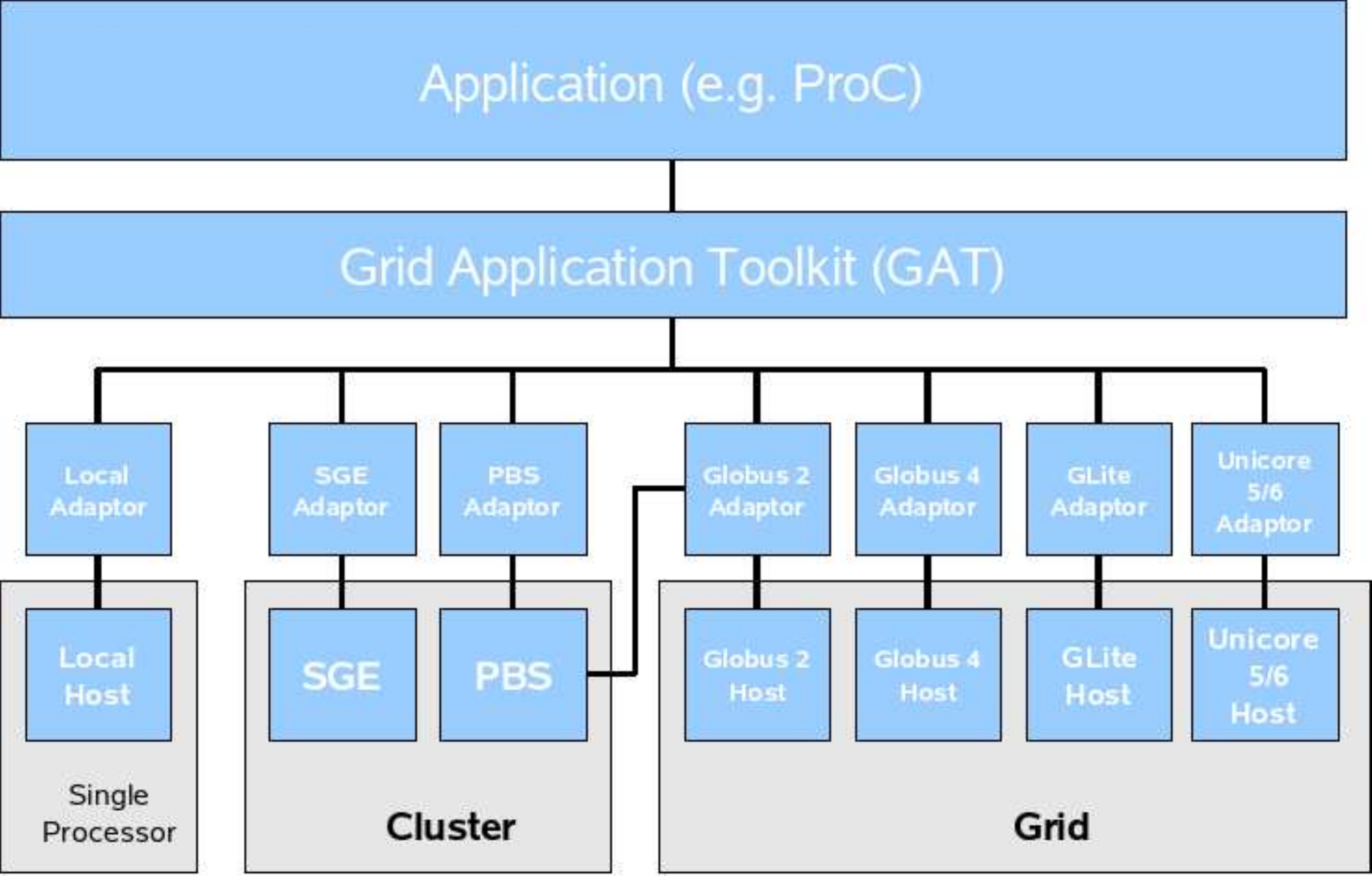}
  \caption{JavaGAT Architecture}
  \label{abb:GATArchitecture}
\end{figure}

The availability of
``local adaptors'' enables the programmer to develop the application
logic without a connection to the grid. The developed application has then 
access to all grid middlewares for which JavaGAT adaptors are available.\\

{\it GridSphere}

Like GAT, {\it (GridSphere, \cite{URL})} was developed 
as part of Gridlab in 2002. The main
goal of the portal related work focused on building a reliable, structured web interface to 
support the European and global grid community.
A portal application can store the specifics of a grid job and run it from any 
standard Web browser.
GridSphere is JSR 168 compliant
and thus portlets running in 
GridSphere can run as well in other portal frameworks.

GridSphere comes with a variety of core portlets providing all the basic 
functionality, such as profile personalisation,
layout customisation and administrative use.

The GridSphere \astrogridd{} portal offers a portlet for
Clusterfinder, and a Cactus Portlet is available at AEI.
\subsection{Components of the \astrogridd{} Architecture}
\label{Components}
The middleware of the \astrogridd{} builds on existing grid tools to
integrate diverse types of resources.
To accommodate the specific requirements of the \astrogridd{} community,
existing components were extended or substituted by newly developed ones.
However, to let other communities benefit from these developments,
we aim at generic solutions wherever possible.

The following subsections describe
(\textit{1.})~the information service Stellaris,
for central storage of all metadata and status information
(\textit{2.})~enhanced data storage capabilities of the grid,
(\textit{3.})~grid access to data sources, efficient data transport
and data streams,
and
(\textit{4.})~options for job submission.

\subsubsection{Information Service}
\label{Inf}
\label{Stellaris}
The goal of the \astrogridd{} information service,
Stellaris~\citep{ges07_stellaris}, is to provide a uniform framework
for storage and querying of grid related information and
metadata. Typical usage scenarios result in questions such as:
\textit{Was data-set X already analysed with program Y and parameter
  set Z? Where is the output data from August 12th last year? Why did
  my last grid job fail? Who created the data producing the graph from
  the latest number of Science and where can I find it?}

Within \astrogridd{}, we distinguish between four different types of
metadata: (1)~\textit{resource metadata} describes properties of the
shared resources (e.g: for a telescope the aperture, filters, ccd, capabilities), (2)~\textit{activity state} reflects the current and
logged state of activities in the grid such as the location and
characteristics of jobs and file transfers (e.g.  user, name of telescope, its location, start and end of observation, priority), (3)~\textit{application
metadata} describes the program and its input parameters (e.g.: RA/Dec of the target, requested filters, etc.), and
(4)~\textit{scientific metadata}, which includes information about the
provenance of data-sets which are used (science project, type of data (image, table), provenance, references, etc.). In order to respond to
the previously stated example questions we will often need to query
metadata of more than one of the information types. Therefore, the
integration of metadata from many different sources is a strong
requirement on the information service. We solve this problem by using
the common metadata model {\it (RDF, \cite{URL})} for all the
information types.

\begin{figure}
  \centering
  \includegraphics[width=\linewidth]{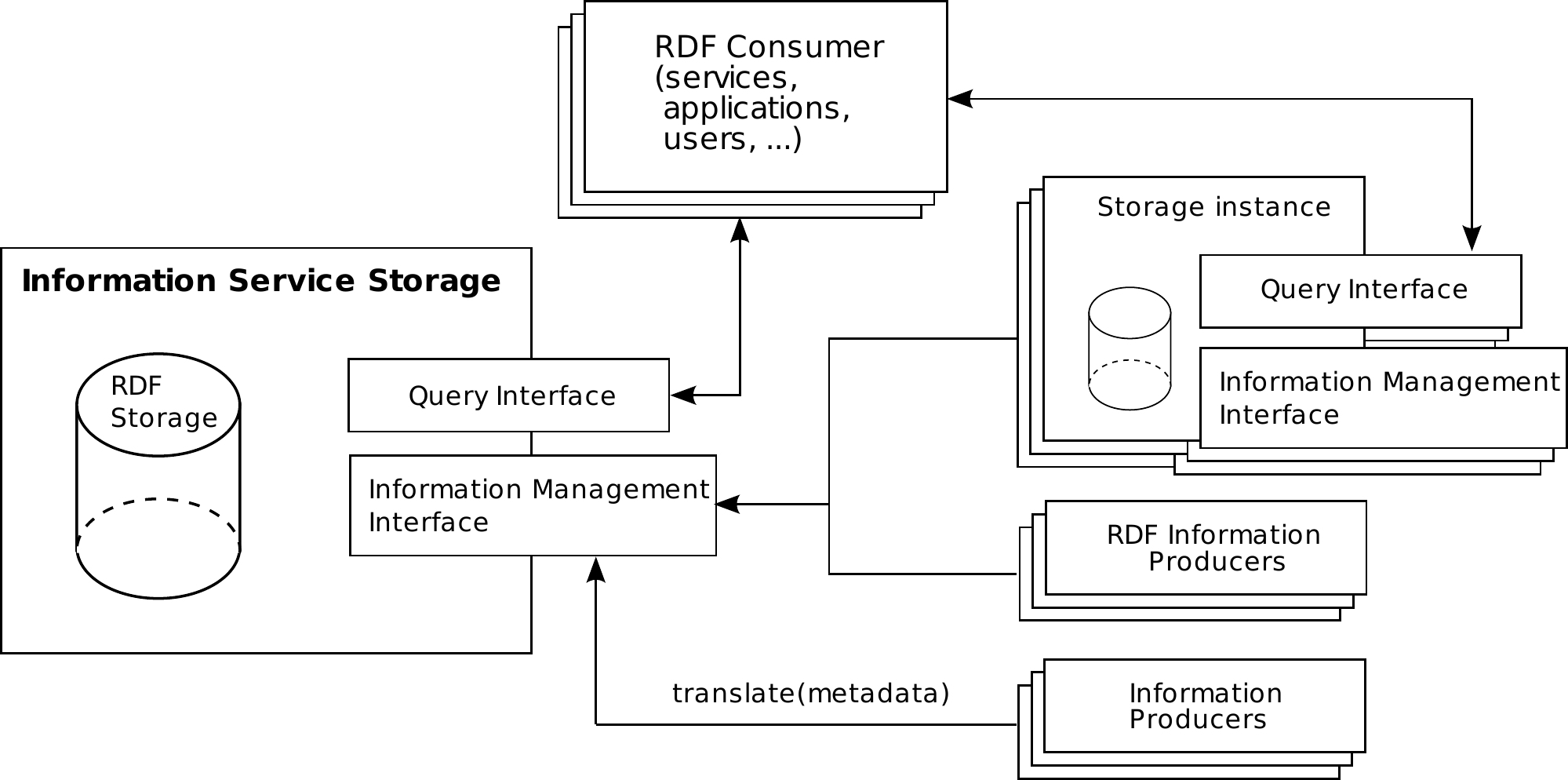}
  \caption{The \astrogridd{} information service framework}
  \label{fig:stellaris-architecture}
\end{figure}

The information system architecture in \astrogridd{} (see
Fig.~\ref{fig:stellaris-architecture}) consists of three main
components; Stellaris, the information service, data producers
(applications, grid resources, and services) and data consumers
(applications, services and users). The Stellaris service itself is
designed around two World Wide Web Consortium (W3C) standards: 
RDF for metadata
representation and {\it (SPARQL, \cite{URL})} which is used for
querying the information service. Thereby, we can benefit from
existing tools for e.g. data integration and visualisation developed
by the web-community at large. The {\it (Stellaris software, \cite{URL})}
was developed within the \astrogridd{} project and is made
available under the Apache Open Source license.
\subsubsection{File Management}
\label{Fil}
\definecolor{sourcecomment}{gray}{0.35}
\newcommand{\sourcecomment}[1]{\textcolor{sourcecomment}{\sourcecode{#1}}}

\newcommand{\sourcecode}[1]{\texttt{#1}}
\newenvironment{sourcecodeENV}{\noindent\begin{quote}\small\ttfamily}{\end{quote
}}%
\newcommand{\globusurlcopy}{\sourcecode{glo\-bus-url-co\-py}}
\newcommand{\globusrlscli}{\sourcecode{glo\-bus-rls-cli}}

The \astrogridd{} Data Management (ADM) has been developed as a tool for 
distributed file management. 
It offers access to the user's files through the concept of a virtual file 
system via the command line, a web interface,
or a programming interface.
Globus contains a software tool denoted as
Globus Replica Location Service (RLS), which allows to manage file
replicas across the grid resources. We found the latter to be somewhat difficult 
to use
with job submission through the GridWay service to an execution
host, whose selection is not directly controlled by the user. 
Our ADM system delivers 
proper software tools to identify files and tag them with metadata independent 
of
the original job execution environment. This is especially useful if the user 
needs to deploy data
files required for job start and to
access files after a job execution for post-processing. 

ADM uses a relational database to store a unique file descriptor, i.e. a logical 
file iden\-ti\-fier
for each file,
plus meta data for each file or directory, e.g. the owner and a timestamp to log 
when
the entry has been registered with the filesystem. 
While file ownership and creation time\-stamp are mandatory, and ADM 
transparently cares for their
maintenance, meta data and individual files can be endowed with custom (user-
defined) properties.
ADM provides the command line client \sourcecode{adm}, including a C-library, 
which offers an easy-to-use access
to the stored
files. Furthermore, ADM ships with a web interface which permits to browse the 
virtual filesystem graphically.

\subsubsection{Data base access and data stream management}
\label{Data}
\begin{figure}
  \centering
  \includegraphics[width=\linewidth]{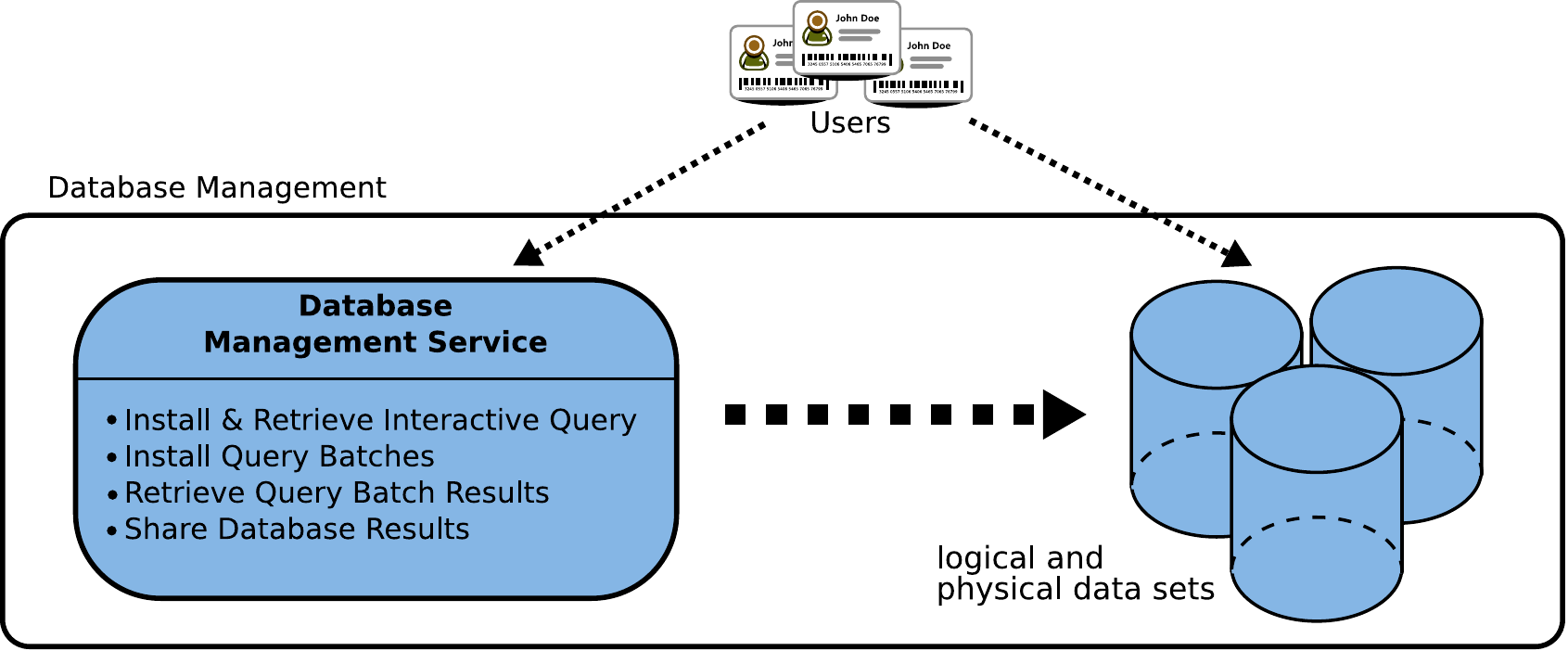}
  \caption{The \astrogridd{} database management. Users can access the
    data sets both interactively and with batch jobs. The actual nodes
    on which the data sets reside is kept transparently.}
  \label{fig:databases}
\end{figure}

Access to \emph{databases} storing observational and simulation data has
become an important part of daily astronomical work. Depending on the
various application requirements and data characteristics, databases
store the actual raw measurements, results and / or the according metadata.
\astrogridd{} considers it a major task to develop database
technology further for building scalable data management
infrastructures. We are motivated by a growing number of users and especially the 
expected data rates of forthcoming projects, such as the
Panoramic Survey Telescope and Rapid Response System (Pan-STARRS)
or LOFAR. 

Due to the distributed nature of data sets and research groups, using a
grid-based approach is a natural choice for the astrophysics community.
The Open Grid Services Architecture---Data Access and
Integration {\it (OGSA-DAI, \cite{URL})} services enable the
integration of databases in grid environments and they are part of the
Globus grid middleware. Therefore we chose OGSA-DAI to provide database
data on resources within the \astrogridd{} and \dgrid{} infrastructure.
Fig.~\ref{fig:databases} gives an overview of the \astrogridd{} database
management.

In order to reduce the network traffic induced by distributed queries on
various data sources and to achieve load balancing within the community
grid, various load balancing techniques have been tested and
evaluated~\citep{hisbase-vldb2007,community-training,hisbase-fgcs2008,workload-aware-hisbase}.

Especially data-centric applications, such as the Clusterfinder use case
(Section~\ref{Clu}), benefit from the increased throughput introduced by
load-balancing techniques for their database accesses (in the case of
Clusterfinder to the SDSS and ROSAT databases). The database relations have
a fixed schema, which is also available via the metadata of the database
system used.  Data access and manipulation is performed via the
standardised query language SQL. In future we also plan to support the 
Virtual Observatory Query Language (VOQL, formerly ADQL, {\it \cite{URL})}), 
a specialised query language for astronomical data based on SQL and an important
effort by the International Virtual Observatory Alliance (IVOA).

\begin{figure}
  \centering
  \includegraphics[width=\linewidth]{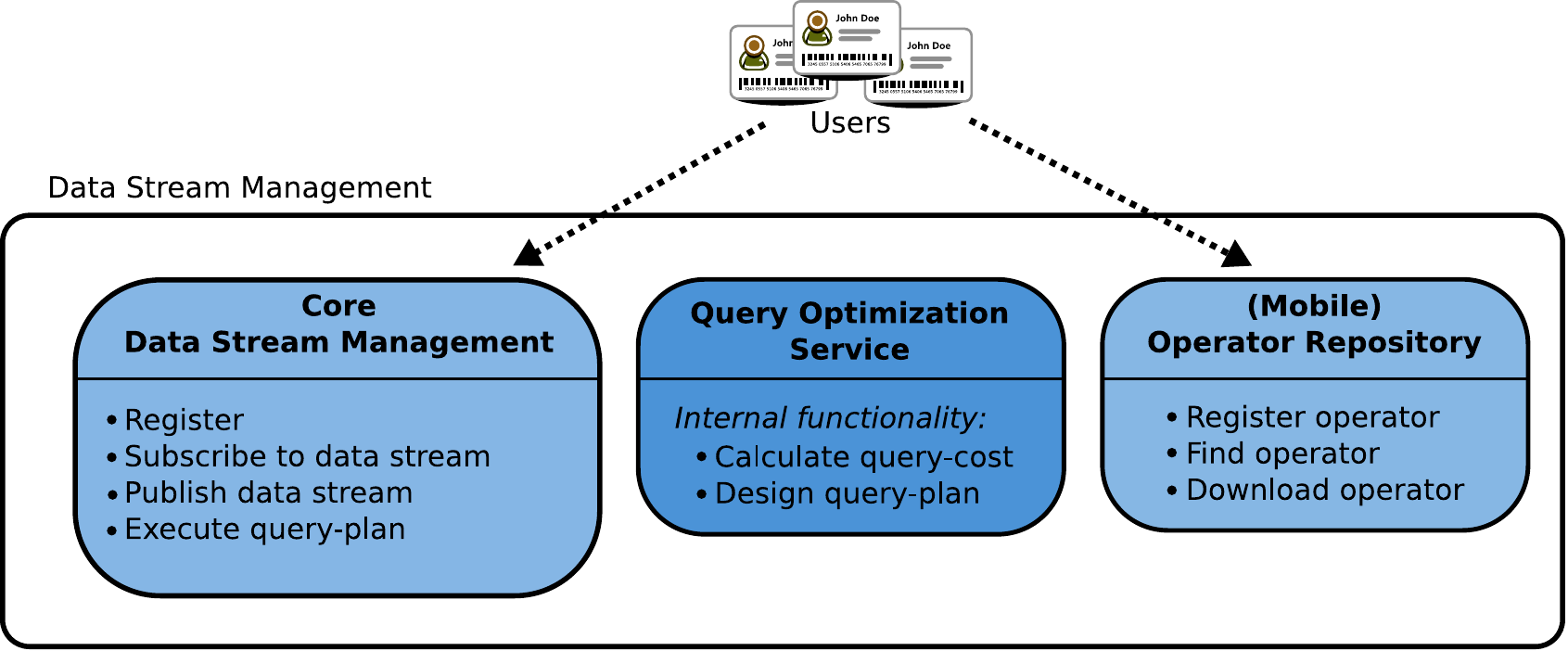}
  \caption{The \astrogridd{} data stream management. Users can publish
    and subscribe to data streams and share their stream-enabled
    operators using operator repositories. Internally, the data stream
    services provide optimisation capabilities. An example of an operator would be a function
	performing a RA/DEC transformation into various coordinate systems. Another example operator 
	would be a Java program listening for specific data from a data stream of an instrument source.}
  \label{fig:datastreams}
\end{figure}

Another prevalent processing model for e-Science data are \emph{data
streams}. Sensor sources (e.g., telescopes, satellites) continuously
generate such data output. Due to the fundamental importance of these
sensors within astrophysics, we investigate efficient data stream
processing models within \astrogridd{}. An important initial processing
step of data streams is data filtering. Existing middleware structures
do not offer such a processing model (yet).

XML or XML-based protocols are the
de-facto communication standard for web services and as well many astronomical
IVOA protocols. Therefore, \astrogridd{} uses XML-based processing
of data streams that are published by data sources and scientific
applications can subscribe to.  In order to increase the reusability of
data streams for multiple subscriptions, the query processing is
performed by installing individual processing steps (\emph{operators})
within the grid network.

Running a data stream management within astrophysics requires means to
define and commonly share scientific operators based on already
implemented functionality. A reusable operator 
is e.g. a chi-squared filter for configurable thresholds for quality assurement.
Mobile operator repositories enable researchers to
provide these operators via their own institution (e.g., personal web
page) and to describe the operators with appropriate metadata in the
information service (Section~\ref{Inf}). This considerably facilitates
collaborating researchers to discover and reuse such existing operators.
Signing the operators with the author's grid certificate allows users to
verify the trustworthiness of the operator's source.

Techniques such as early filtering and early aggregation lead to good
results, especially in the context of multi-subscription
optimisation~\citep{KuntschkeSKR-VLDB05DEMO,KuntschkeK-LNCS2006,KuntschkeK-CIKM2006}.
The \astrogridd{} data stream management (see
Fig.~\ref{fig:datastreams}) is available on all \astrogridd{} resources.

By developing data stream processing techniques for grid environments, we
moreover support the conversion from persistent data sets to streams. A
combined, integrated processing of persistent and streaming data, as
required by applications such as SED classification, is
possible and results in better performance~\citep{starglobe}.
\subsubsection{Job Management}
\label{Job}
\label{Gridway}

\astrogridd{} has implemented job management through the independently developed
{\it (GridWay \cite{URL})} Metascheduler
on top of the standard globus middleware layer. As a metascheduler, GridWay
enables large-scale, reliable and efficient sharing of computing resources
managed by different Local Resource Management (LRM) systems, such as the
Portable Batch System (PBS), the Sun Grid Engine (SGE), or the
LSF,
within a single organisation (enterprise grid) or scattered across several 
administrative 
domains. In the second case GridWay can interact also with other grid middleware 
than Globus, such as e.g. Unicore or gLite. 
GridWay is meanwhile fully
integrated into the globus open source project, adheres to Globus philosophy and
guidelines for collaborative development and so welcomes code and support 
contributions.
GridWay has its own set of line mode commands, such as e.g. gwsubmit, gwstat
or gwhosts to control the available resources and one's own jobs. 
GridWay can serve as a comfortable user interface to the entire grid,
similar in style to a local resource management system (LRM, queue system).
Note that resource informations have to be provided through the Globus MDS 
information
service and middleware to the GridWay server. 
The LRM ``Fork'' means that single processor jobs are accepted to be 
started by
a Unix process fork. Another LRM available is PBS (portable batch system) for
parallel jobs.
\begin{figure}
  \centering
  \includegraphics[width=\linewidth]{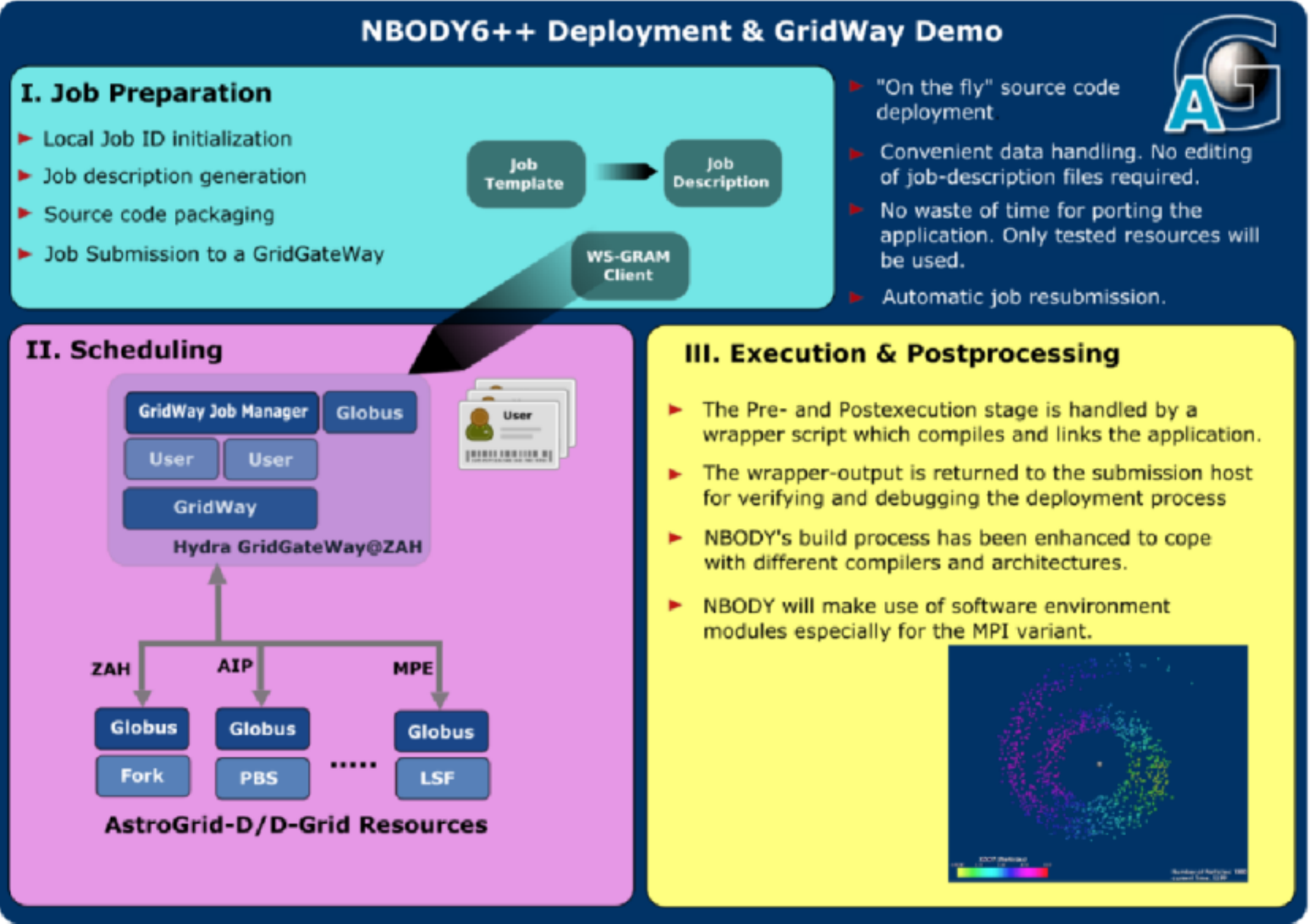}
  \caption{Flowchart of Steps to Submit NBODY Job via GridWay}
  \label{abb:GridWayNBODY}
\end{figure}
Fig.~\ref{abb:GridWayNBODY} illustrates three stages of a job run, 
for the example of an NBODY calculation (\ref{Nbo}). The first step is the
deployment which delivers an XML based job description as described in
Section~\ref{Nbo}. Such XML jobs can be submitted through the standard Globus 
GRAM job submission interface and middleware to the Gridway host rather than directly
to the LRM of an execution host. 

Gridway then receives this job through Globus and
the Gridway Job Manager acts as a broker and scheduler. It selects an 
available execution host through a matchmaking process and submits the job 
to it by Globus GRAM. At present we have 
implemented a simple round robin strategy for single fork jobs; the GridWay 
software in principle allows to implement more complex scheduling algorithms 
including user defined parameters. It is always possible to submit jobs targeted to a 
certain resource through GridWay, though this is not the desired mode of operation. 

The third step is the execution and postprocessing stage, during which it has 
to be ensured that the build process properly works on the target resource and 
that the user receives the simulation results for postprocessing. 

The two-
step submission procedure with two Globus GRAM jobs connected by the GridWay server is
denoted as \emph{GridGateWay}. Note that it is also possible for the user to directly 
logon to the GridWay host and use it for job submission directly.

\section{Summary and Outlook}
\label{Con} \label{Summary}
\subsection*{Summary}
\astrogridd{} established a nation-wide pool of compute, data, 
and instrument resources accessible for astronomers. 
It also integrated special hardware compute resources 
like clusters of GRAPE6 boards into the grid. The use case NBODY6++ shows impressively it's 
exploit in a grid environment. 
Well documented procedures explaining how to bring a resource into 
the grid are available. Authentication and authorisation for the use of the grid resources is managed by the
Virtual Organisation. Moreover, the resources of 
\astrogridd{} were integrated in \dgrid{}, which in turn provides access to 
the resources of the whole \dgrid{} for the \astrogridd{} members. 
Robotic telescopes were 
also integrated into the grid as a special hardware resource, so they can be accessed like any other compute node.

A variety of typical astronomical applications was brought to the grid. 
We investigated simple but compute intensive 
task farming applications like Dynamo or GEO600 and showed that it is very easy to
run them on the grid without the need of complex reprogramming. We also 
looked into more complex and data intensive tasks like e.g. the Clusterfinder and
ported them to the grid.  
The Clusterfinder program, e.g., is now able to scan the
entire available data for one model parameter set within several days, whereas it would
need more than two years on a single processor. 

We developed a 
set of high-level services: Programmers can now make use of an information
service to handle meta data and to monitor jobs and resources.
Also, they can abstract from interfacing a specific grid middleware and use GAT instead. Moreover, 
GridSphere enables a user friendly grid access with any 
web browser. The ProC workflow engine supports the composition of 
scientific workflows and their parallel grid execution. 
Resource brokering and job scheduling is augmented in \astrogridd{} by the
GridWay Metascheduler. Thereby more complex scheduling algorithms can be 
implemented.
The \astrogridd{} Data Management ADM handles file staging 
in combination with the job submission via GridWay.
It thus provides an easy-to-use access to stored files and their replica in the grid.
The integration of databases and data streams is also provided by 
\astrogridd{}. Special attention is paid to optimising techniques that 
guarantee good performance results as well for throughput as for response time. 
Many of the services summarised above are addressed in
close collaboration with GAVO, whose focus is more on the side of the 
scientific user, whereas
\astrogridd{} is solving the technical and infrastructural aspects.

Most of the German community grids, except the High Energy Physics community,
employ the Globus Middleware. 
On EU level, gLite developed by EGEE is dominating all grid efforts, whereas internationally, the split 
is equal between EGEE/gLite and Globus. 
A lot of effort goes into interoperability of these 
different middlewares, but sometimes there still are barriers. 
\astrogridd{} is collaborating with both EGEE/EGI
as well as the Open Science Grid (OSG).

\subsection*{Outlook}
The important next step is to enlarge the community of grid users. For this 
purpose, the consulting 
and the support of new users has to be professionalised. We are
able to offer considerable resources in compute power and storage to the
scientific community.

There are some infrastructure elements that we would like to improve, e.g.
our methods for resource brokering and 
job scheduling. 

Proper and efficient handling of large amounts of data is a key
feature that the grid offers. Upcoming projects 
such as LOFAR, PanStarrs or LSST will produce immense data 
volumes whose storage, administration, and processing can no
longer be handled by local institutions. Moreover, this data is in many cases 
processed in distributed, international working groups.
Grid technology is an appropriate answer to these new challenges. Due to the 
parallelisation potential and the security layers of the grid, administration and access can be achieved 
even in a complexity where central processing hits its limit. 
For this purpose we need a 
powerful data management component to enable handling files, data
bases, and data streams in a coherent framework. 

\astrogridd{} established a solid basis to cope with these future 
challenges arising from forthcoming scientific needs. We are looking
forward to establish our solutions as a cornerstone of German
e-Astronomy.

\ 

\noindent 
{\bf Acknowledgments}\\
This work is supported by the German Federal Ministry of Education and
Research within the \dgrid{} initiative under contracts 01AK804[A-G].
AIP acknowledges support by EFRE, grant No. 9053 
ARI-ZAH acknowledges support of the GRACE project by Volkswagen Foundation
grant No. I/80\,041-043 (Project {\sc 'GRACE'}) and by 
the Ministry of Science, Research and the Arts of Baden-W\"urttemberg (Az: 
823.219-439/30 and /36).
We acknowledge the special memorandum of
understanding between Astrogrid-D and the astronomical segment of
Ukrainian Academic GRID Network.
We thank Ignacio Llorente, Ruben Montero, and Tino V{\'a}zquez of Universidad 
Complutense
Madrid, Spain, for help and support in installation and operation of the GridWay 
service.
\appendix

\bibliographystyle{elsarticle-harv}
\bibliography{references}

\end{document}